\documentstyle[referee,graphicx]{mn}

\newif\ifAMStwofonts

\ifoldfss
  \ifCUPmtlplainloaded \else
    \NewTextAlphabet{textbfit} {cmbxti10} {}
    \NewTextAlphabet{textbfss} {cmssbx10} {}
    \NewMathAlphabet{mathbfit} {cmbxti10} {} 
    \NewMathAlphabet{mathbfss} {cmssbx10} {} 
  \fi
  \ifAMStwofonts
    \ifCUPmtlplainloaded \else
      \NewSymbolFont{upmath} {eurm10}
      \NewSymbolFont{AMSa} {msam10}
      \NewMathSymbol{\upi}     {0}{upmath}{19}
      \NewMathSymbol{\umu}     {0}{upmath}{16}
      \NewMathSymbol{\upartial}{0}{upmath}{40}
      \NewMathSymbol{\leqslant}{3}{AMSa}{36}
      \NewMathSymbol{\geqslant}{3}{AMSa}{3E}
      \let\oldle=3D\le     \let\oldleq=3D\leq
      \let\oldge=3D\ge     \let\oldgeq=3D\geq
      \let\leq=3D\leqslant \let\le=3D\leqslant
      \let\geq=3D\geqslant \let\ge=3D\geqslant
    \fi
  \fi
\fi 

\ifnfssone
  \newmathalphabet{\mathit}
  \addtoversion{normal}{\mathit}{cmr}{m}{it}
  \addtoversion{bold}{\mathit}{cmr}{bx}{it}
  \newmathalphabet{\mathbfit} 
  \addtoversion{normal}{\mathbfit}{cmr}{bx}{it}
  \addtoversion{bold}{\mathbfit}{cmr}{bx}{it}
  \newmathalphabet{\mathbfss} 
  \addtoversion{normal}{\mathbfss}{cmss}{bx}{n}
  \addtoversion{bold}{\mathbfss}{cmss}{bx}{n}
  \ifAMStwofonts
    \ifCUPmtlplainloaded \else
      \UseAMStwoboldmath
      \makeatletter
      \new@mathgroup\upmath@group
      \define@mathgroup\mv@normal\upmath@group{eur}{m}{n}
      \define@mathgroup\mv@bold\upmath@group{eur}{b}{n}
      \edef\UPM{\hexnumber\upmath@group}
      \new@mathgroup\amsa@group
      \define@mathgroup\mv@normal\amsa@group{msa}{m}{n}
      \define@mathgroup\mv@bold\amsa@group{msa}{m}{n}
      \edef\AMSa{\hexnumber\amsa@group}
      \makeatother
      \mathchardef\upi=3D"0\UPM19
      \mathchardef\umu=3D"0\UPM16
      \mathchardef\upartial=3D"0\UPM40
      \mathchardef\leqslant=3D"3\AMSa36
      \mathchardef\geqslant=3D"3\AMSa3E
      \let\oldle=3D\le     \let\oldleq=3D\leq
      \let\oldge=3D\ge     \let\oldgeq=3D\geq
      \let\leq=3D\leqslant \let\le=3D\leqslant
      \let\geq=3D\geqslant \let\ge=3D\geqslant
    \fi
  \fi
\fi 

\ifnfsstwo
  \DeclareMathAlphabet{\mathbfit}{OT1}{cmr}{bx}{it}
  \SetMathAlphabet\mathbfit{bold}{OT1}{cmr}{bx}{it}
  \DeclareMathAlphabet{\mathbfss}{OT1}{cmss}{bx}{n}
  \SetMathAlphabet\mathbfss{bold}{OT1}{cmss}{bx}{n}
  \ifAMStwofonts
    \ifCUPmtlplainloaded \else
      \DeclareSymbolFont{UPM}{U}{eur}{m}{n}
      \SetSymbolFont{UPM}{bold}{U}{eur}{b}{n}
      \DeclareSymbolFont{AMSa}{U}{msa}{m}{n}
      \DeclareMathSymbol{\upi}{0}{UPM}{"19}
      \DeclareMathSymbol{\umu}{0}{UPM}{"16}
      \DeclareMathSymbol{\upartial}{0}{UPM}{"40}
      \DeclareMathSymbol{\leqslant}{3}{AMSa}{"36}
      \DeclareMathSymbol{\geqslant}{3}{AMSa}{"3E}
      \let\oldle=3D\le     \let\oldleq=3D\leq
      \let\oldge=3D\ge     \let\oldgeq=3D\geq
      \let\leq=3D\leqslant \let\le=3D\leqslant
      \let\geq=3D\geqslant \let\ge=3D\geqslant
    \fi
  \fi
\fi 

\ifCUPmtlplainloaded \else
  \ifAMStwofonts \else 
    \def\upi{\pi}
    \def\umu{\mu}
    \def\upartial{\partial}
  \fi
\fi

\newcommand{\lsim}{\hbox{ \rlap{\raise 0.425ex\hbox{$<$}}\lower 0.65ex\hbox{$\sim$} }}
\newcommand{\gsim}{\hbox{ \rlap{\raise 0.425ex\hbox{$>$}}\lower 0.65ex\hbox{$\sim$} }}

\title[Massive Star Populations in Wolf-Rayet Galaxies]{Massive Star Populations in Wolf-Rayet Galaxies}
\author[I. F. Fernandes et al.]{I. F. Fernandes$^{1}$, R.
de Carvalho$^{2}$, T. Contini$^{3}$ and R. R. Gal$^{4}$ \\
$^{1}$Instituto Astron\^{o}mico e Geof\'{i}sico - USP, Rua do
Mat\~{a}o 1226, CEP 05508-900, S\~{a}o Paulo, Brazil\\
E-mail:iran@astro.iag.usp.br\\
$^{2}$INPE / DAS,  Av. dos Astronautas, 1.758, CEP 12227-010,
S\~{a}o Jos\'{e} dos Campos,  Brazil\\
E-mail:reinaldo@das.inpe.br\\
$^{3}$Laboratoire d'Astrophysique (UMR 5572), Observatoire
Midi-Pyr\'{e}n\'{e}es, 14 Avenue Edouard Belin,F-31400, Toulouse,
France\\
E-mail:contini@ast-omp.fr\\
$^{4}$Department of Physics, UC Davis, One Shields Ave., Davis, CA
95616, USA\\
E-mail:gal@physics.ucdavis.edu}

\begin{document}

\date{Accepted        .     Received        .}

\pagerange{\pageref{firstpage}--\pageref{lastpage}} \pubyear{2004}

\maketitle \label{firstpage}

\begin{abstract}
We analyze longslit spectral observations of fourteen Wolf-Rayet
galaxies from the sample of Schaerer, Contini \& Pindao (1999).
All 14 galaxies show broad Wolf-Rayet emission in the blue region
of the spectrum, consisting of a blend of NIII$ \lambda4640$,
CIII$\lambda4650$, CIV$\lambda4658$, and HeII$ \lambda4686$
emission lines, which is a spectral characteristic of WN stars.
Broad CIV$\lambda5808$ emission, termed the red bump, is detected
in 9 galaxies and CIII$\lambda5996$ is detected in 6 galaxies.
These emission features are due to WC stars. We derive the numbers
of late WN and early WC stars from the luminosity of the blue and
red bumps, respectively. The number of O stars is estimated from
the luminosity of the H$\beta$ emission line, after subtracting
the contribution of WR stars. The Schaerer {\&} Vacca 1998
(hereafter SV98) models predict that the number of WR stars
relative to O stars, $N_{WR}/N_{O}$, increases with metallicity.
For low metallicity galaxies, the results agree with predictions
of evolutionary synthesis models for galaxies with a burst of star
formation, and indicates an IMF slope -2$\lsim\Gamma\lsim-2.35$ in
the low metallicity regime. For high metallicity galaxies our
observations suggest a Salpeter IMF ($\Gamma = -2.35$) and an
extended short burst.  The main possible sources of error are the
adopted luminosities for single WCE and WNL stars. We also report,
for the first time, NGC 450 as a galaxy with WR characteristics.
For NGC 450, we estimate the number of WN and WC stars. The number
ratio $N_{WR}/N_{O}$, and the equivalent widths of the blue bump,
$EW_{\lambda4686}$, and of the red bump, $EW_{\lambda5808}$ in NGC
450 are also in good agreement with the instantaneous burst model
prediction for WR galaxies.
\end{abstract}


\section{Introduction}

Wolf-Rayet galaxies are extragalactic objects whose spectra show
direct signatures similar to those observed in Wolf-Rayet (WR)
stars. The most common characteristic is the presence of a broad
HeII$\lambda4686$ feature (the blue bump) originating in the
stellar winds of WR stars (Schaerer Contini \& Pindao 1999,
hereafter SCP99). WR galaxies have long been known, with the first
discovery of such spectral features in the blue compact galaxy He
2-10 (Allen, Wright \& Goss 1976). The concept of WR galaxies was
introduced by Osterbrock \& Cohen (1982, hereafter OC82) and Conti
(1991).

The blue WR bump is often blended with nearby nebular emission
lines of He, Fe, or Ar, and can show several broad stellar
emission components (NIII$\lambda4640$, CIII$\lambda4650$,
HeII$\lambda4686$) which are difficult to deblend in most low- or
medium-resolution spectra. These features originate in WR stars of
WN and/or WC subtypes (OC82 and Conti 1991). The strongest
emission line in WC stars is CIV$\lambda5808$, which is very weak
in WN stars. This ``red W-R bump" has only rarely been observed.
Where the data is available, CIV$\lambda5808$ is generally weaker
than HeII$\lambda4686$.

In more distant galaxies, WR stars can only be indirectly
detected, by observing the integrated spectra of the galaxies.
Strong star formation activity indicates the presence of a large
number of massive stars, most of which evolve through the WR
phase. At a given stage of the starburst, many WR stars appear,
but only for a brief duration. Thus, the presence of WR features
in these galaxies indicates recent star formation ($<$10 Myr) as
well as the presence of massive stars ($M_{initial} >
25M_{\odot}$) (Schaerer et al. 1999). This provides interesting
constraints on recent star formation episodes in these objects
(Maeder \& Conti 1994). Furthermore, metallicity plays an
important role in regulating the lower mass limit above which a
star passes through the WR phase.

Despite their small number compared to other massive stars,
especially in low-metallicity galaxies, WR stars are numerous
enough for their integrated emission to be detected. In this
paper, we shall follow OC82 and Conti (1991): A WR galaxy is
classified as such if its integrated spectrum shows detectable WR
broad features emitted by unresolved stellar clusters.

The compilation of Conti (1991) included only 37 objects. Since
then, the number of cataloged WR galaxies has increased rapidly,
with more than 130 known today (Guseva, Izotov \& Thuan 2000,
hereafter GIT00; SCP99). WR galaxies do not form a homogeneous
class, exhibiting a variety of morphologies. Among the WR galaxies
we find low-mass blue compact dwarf (BCDs), irregular galaxies,
massive spirals and ultra-luminous merging IRAS galaxies. Recent
studies show that WR features are also seen in LINERs and Seyfert
2 galaxies (OC82; Ho et al. 1995; Heckman et al. 1997; Schmitt et
al. 1998 and Contini et al. 2001). The possibility of detecting WR
stars in central cluster galaxies out to a redshift of $z\geq0.25$
is discussed in Allen (1995).

The number of WR stars relative to massive stars is highly
dependent on metallicity. Theoretical evolutionary models predict
that at fixed metallicity, the ratio between WR and other massive
stars varies strongly with the age of the starburst (Mass-Hesse \&
Kunth 1991; Maeder 1991; Maeder \& Meynet 1994; Meynet 1995;
SV98). The maximum value of this ratio decreases from 1 to 0.02
when the metallicity decreases from $Z_{\odot}$ to $Z_{\odot}/50$
(GIT00). Similarly, the duration of the WR stage in the starburst
also decreases with decreasing metallicity. Hence, the number of
galaxies with extremely low metallicity containing WR stellar
populations is expected to be small.

GIT00 derived the number of WCE and WNL stars from the luminosity
of the red and blue bumps, respectively, and the number of O stars
from the H$\beta$ luminosity, for 39 WR galaxies with heavy
element mass fractions between 1/50 and twice solar. In their
sample, the blue bump consists of an unresolved blend of WR and
nebular lines. They proposed a new technique to derive the number
of WNL stars using NIII$\lambda4512$ and SiIII$\lambda4565$
emission lines. They found that the relative number of Wolf-Rayet
stars N$_{WR}$/N$_{(O+ WR)}$ and N$_{WC}$/N$_{WN}$ derived from
observations are in satisfactory agreement with theoretical
predictions (SV98). The results obtained for the extremely
metal-poor galaxies disagree with model predictions likely due to
the low emission line luminosity for WCE stars in metal-poor
models.

Schaerer et al. (2000, hereafter SGIT00), using five metal rich
objects from GIT00 and new results on Mrk 309, attempted to
constrain the properties of massive star populations and star
formation histories by comparing their observations with
evolutionary synthesis models. They found that extended burst
durations of $\sim4-10$ Myr or a superposition of several bursts
were required to produce the observed WR population and red
supergiant features. The burst durations are longer than those
obtained for other objects in Schaerer et al. (1999) using the
same models.

Pindao et al. (2002) analyzed the spectra of 85 high metallicity
disk HII regions of nearby spiral galaxies. In contrast with
previous studies of low metallicity galaxies, they found smaller
values of I(WR)/I(H$\beta$) than predicted from evolutionary
models at corresponding metallicities. They suggested the use of
two WR luminosity regimes to correct the model predictions.

The goal of our study of these 14 galaxies is to search for and
confirm the presence of WN and WC stars in galaxies with different
metallicities, and to compare the results obtained for this sample
with predictions from evolutionary synthesis models (SV98) and
starbust99 (Leitherer et al. 1999). We also report NGC 450 as a
newly classified WR galaxy.

The paper is structured as follows: the observations and
procedures used to reduce the data are described in \S2. In this
section we also discuss corrections for reddening and underlying
absorption affecting the emission lines. Section 3 describes how
contamination of a starburst spectrum by the presence of SN IIe
and AGN is accounted for, while Section 4 explains how the
physical parameters of the gas are calculated. The massive star
population and constraints on the evolutionary tracks of the
starburst regions (age, burst duration, and IMF) are derived in
\S5 from a comparison with evolutionary synthesis models. Finally,
our main results are summarized and discussed in \S6.

\section{Spectroscopic Observations and Data Reduction}

\subsection{Observations}

We observed 14 galaxies from the sample of Wolf-Rayet galaxies and
extragalactic HII regions presented in SCP99. NGC 6764 was
observed at two position angles, $67^{\circ}$ and $90^{\circ}$,
with the former along the major axis of the galaxy and the latter
spanning a secondary emission region near the central region. We
suspected that WR stars may be present in this region, but the
spectra do not confirm this hypothesis.

Data were gathered at two different sites. We used the Palomar
200-inch telescope with the longslit Double Spectrograph (Oke \& Gunn
1982) on UT 1999 October 10-11, with the 1200 l/mm grating, blazed at
$5000$\AA, yielding a pixel size of $0.62^{\prime\prime}\times1.3$\AA~
in the blue and $0.47^{\prime\prime}\times1.7$\AA~ in the red, and a
total wavelength coverage of $3600-6700$\AA. The slit width was set to
$1^{\prime\prime}$, resulting in a spectral resolution of
$\sim5.6$\AA~ in the blue and $\sim5.7$\AA~ in the red. The slit
length was 180$^{\prime\prime}$. Additional data were taken at the
$3.6m$ ESO/NTT telescope using the ESO Multi-Mode Instrument (EMMI)
with CCD36 on UT 1999 April 17. The total wavelength coverage was
4000-6600\AA. The slit width was set to $1^{\prime\prime}$, resulting
in a spectral resolution of $\sim5.9$\AA. The length of the slit was
$120^{\prime\prime}$.

The slit was aligned along the major axis of the galaxy when
possible, and centered on the brightest region of the target. The
journal of observations of all objects in the sample is provided
in Table 1. Figures 1(a)-(o) show the images of each galaxy taken
from the DSS with the slit position overlaid.

\setcounter{figure}{0}
\begin{figure}

{(a)}\includegraphics[width=75mm]{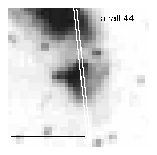}
{(b)}\includegraphics[width=75mm]{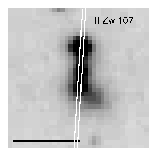}
{(c)}\includegraphics[width=75mm]{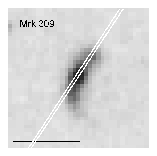}
{(d)}\includegraphics[width=75mm]{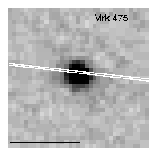}

\end{figure}

\setcounter{figure}{0}
\begin{figure}

{(e)}\includegraphics[width=75mm]{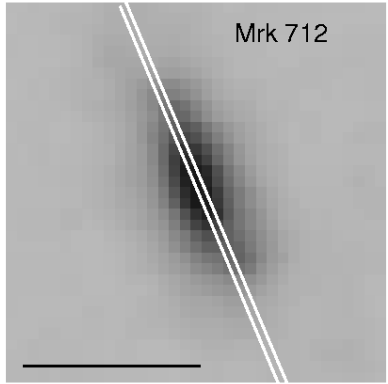}
{(f)}\includegraphics[width=75mm]{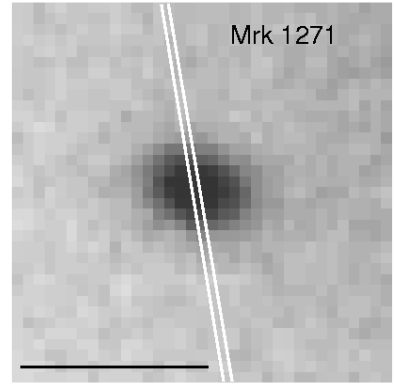}
{(g)}\includegraphics[width=75mm]{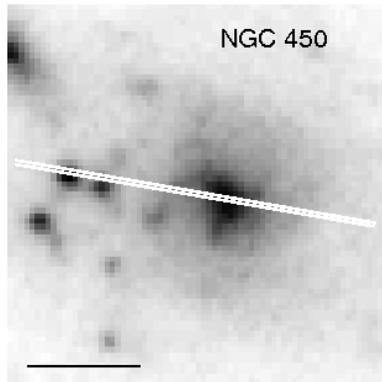}
{(h)}\includegraphics[width=75mm]{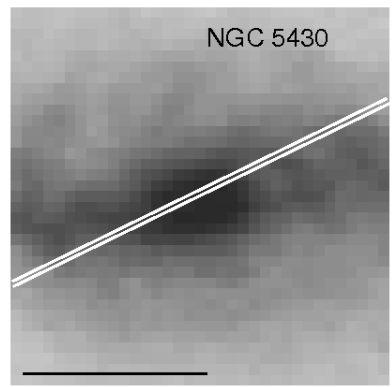}

\caption{DSS images of galaxies with slit positions. (a) Fairall
44, (b) III Zw 107, (c) Mrk 309, (d)Mrk 475, (e) Mrk 712, (f) Mrk
1271, (g) NGC 450, (h) NGC 4385. The horizontal bar at the lower
left indicates 0.5 arcminutes.} \label{fig1}

\end{figure}

\setcounter{figure}{0}
\begin{figure}

{(i)}\includegraphics[width=75mm]{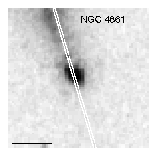}
{(j)}\includegraphics[width=75mm]{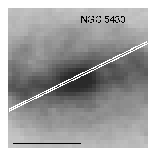}
{(k)}\includegraphics[width=75mm]{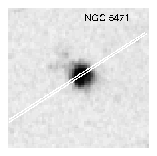}
{(l)}\includegraphics[width=75mm]{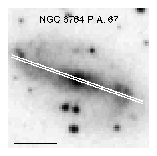}

\end{figure}

\setcounter{figure}{0}
\begin{figure}

{(m)}\includegraphics[width=75mm]{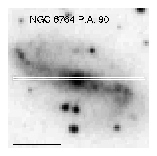}
{(n)}\includegraphics[width=75mm]{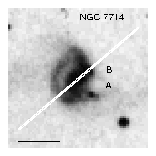}
{(o)}\includegraphics[width=75mm]{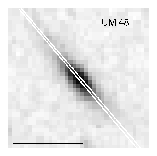}

\caption{Continued. DSS images of galaxies with slit positions. (i)
NGC 4861,(j) NGC 5430, (k) NGC 5471, (l) NGC 6764 P.A. 67, (m) NGC
6764 P.A. 90, (n) NGC 7714 and (o) UM48. The horizontal bar at the
lower left indicates 0.5 arcminutes.} \label{fig1}
\end{figure}

\subsection{Data Reduction}

Initial data reduction was carried out following standard
procedures with the IRAF task $ccdproc$. The package includes bias
subtraction, flat-field correction, subtraction of the night sky
background and bad column removal. The bias level was subtracted
from each frame using the overscan region of the CCD chip.

Spectrum extraction was performed using the IRAF task $longslit$.
The task includes correction for atmospheric extinction,
wavelength calibration and flux calibration. Cosmic rays were
removed using the $cosmicrays$ task with a threshold of $5\%$.

Spectrophotometric standard stars were observed each night to
perform flux calibration. Arc lamps were taken before and after
each exposure in order to provide accurate wavelength calibration.
An average rms of 0.1\AA~ was obtained for the pixel to wavelength
fit using a third order spline.

We use a standard extraction aperture for each object whose width
is such that the peak intensity of the H$\beta$ line decreases by
$80\%$ from the center, along the spatial direction. This width is
set independently for each side of the peak. For objects with more
than one emission region, different apertures were set for each
knot. With this size we obtain the necessary spectral properties
for this study while minimizing possible contamination from the
adjacent stellar population.

The signal to noise ratio, S/N, was determined assuming Poisson
statistics and using the readout noise and gain of the CCD, the
number of combined spectra, and the sky value. For NTT data the
readout noise is 5.43e$^{-1}$ with a gain of 2.18 e$^{-1}$ per
ADU. For the Palomar observations, the readout noise is 8.6
e$^{-1}$ with a gain of 2.13 e$^{-1}$ per ADU in the blue channel
and 7.5 e$^{-1}$ and 2.00 e$^{-1}$ per ADU in the red channel. The
values of S/N are given in Table 1 for the NTT, as well as the
blue and red parts of the Palomar spectra, where they are averages
over a range of 100 \AA~ around 4686\AA~ and 5878\AA,
respectively.

\setcounter{table}{0}
\begin{table}
\begin{minipage}{220mm}
\caption{Journal of Observations}
\begin{tabular}{lrrccrllrrl}
\hline \hline

Galaxy& $\alpha$& $\delta$& Exp. Time& S/N& P.A.&
Observat.& Type& D& Scale& mag\\
& ($\fh\fh$:$\fm\fm$:$\fs\fs$)& ($\degr$:$\arcmin$:$\arcsec$)&
($s$)& (\AA$^{-1}$)& ($\degr$)&
& &($Mpc$)& (pc/$\arcsec$)& (Band)\\
\hline

UM 48& 0:36:10& 4:38:06&            $3\times$  1200&       96/73& 43&    Palomar&          S&      65.22&        316& 13.35(J)\\

NGC 450& 1:15:31& -0:51:38&        1500 + 1200&            90/77&  80&    Palomar&   SAB(s)cd&      22.46&        109& 11.40(J)\\

MRK 712& 9:56:42& 15:38:04&         $4\times$ 900&         84& 108 &        NTT&       SBbc&      55.87&        271& 14.47(B)\\

MRK 1271& 10:56:08& 6:10:15&        $4\times$ 900&         80&  98&        NTT&    Compact&       8.79&         43& 14.80(mP)\\

NGC 4385& 12:25:42& 0:34:23&        $4\times$ 900&         96& 192&        NTT&   SB(rs)0+&      23.35&        113& 10.64(J)\\

NGC 4861& 12:59:00& 34:50:43&       1800 + 600&            160/97&  19&    Palomar&     SB(s)m&      10.93&         53& 12.44(J)\\

NGC 5430& 14:00:46& 59:19:35&        $2\times$ 1200&       140/104& 146&    Palomar&     SB(s)b&      40.55&        197& 10.01(J)\\

NGC 5471& 14:04:29& 54:23:49&       $2\times$ 1200&        91/75& 127&    Palomar&        HII&       3.86&         19& 14.76(J)\\

MRK 475& 14:39:05& 23:37:28&        $2\times$ 1800&        84/70&  80&    Palomar&        BCD&       7.96&         39& 15.46(B)\\

Fairall 44& 18:13:38& -57:43:59&      $2\times$ 900&       90&        105&        NTT&     S? pec&      63.62&        308& 12.34(J)\\

NGC 6764& 19:08:15& 50:55:57&       $2\times$ 1200&         125/95& 67/90&    Palomar&    SB(s)bc&      31.26&        152& 10.56(J)\\

MRK 309& 22:52:34& 24:43:50&       $2\times$ 1500 + 1800&   123/90&  140&    Palomar&         Sa&     167.09&        810& 13.05(J)\\

III ZW 107& 23:30:09& 25:32:02&    1800 + 1200&           125/95&      175&    Palomar&         Sb&      76.50&        371& 14.06(J)\\

NGC 7714& 23:36:13& 2:09:21&        $2\times$  1200&        99/71& 122&    Palomar& SB(s)b:pec&      37.05&        180& 10.77(J)\\
\hline \hline
\medskip
\end{tabular}
\end{minipage}
($\alpha$) and ($\delta$) coordinates (J2000) are for the slit
center. The distance (D) is obtained from the redshift measured
directly from the object spectrum using H$_0$ =75 km s$^{-1}$
Mpc$^{-1}$. NTT and Palomar indicate the observatory where the object
was observed. The object type is obtained from NED. The linear scale
is obtained using the computed distance, D. $\ast$For more details on
the quoted magnitudes we refer the reader to the NED-IPAC database.
\end{table}

\subsubsection{Reddening Correction and Underlying Balmer Absorption}

The reddening correction is obtained from the Balmer line ratios,
using the extinction law of Cardelli, Clayton $\&$ Mathis (1989)
(hereafter CCM89), and the theoretical Balmer emission line ratios for
case B recombination: $I(H\alpha)/I(H\beta)$ = 2.88 and
$I(H\gamma)/I(H\beta)$ = 0.47 (Brocklehurst 1971). We obtain the color
excess, $E(B - V)$, using the theoretical ratio of Balmer to
$F(H\alpha)/F(H\beta)$ measurements, according to

\begin{equation}
E(B - V) = 2.33 \times \left[ {Log\left( {\frac{{F(H\alpha
)}}{{F(H\beta )}}} \right) - Log(2.88)} \right]
\end{equation}

In three cases (Mrk 309, Mrk 1271, and Fairall 44) for which
$F(H\alpha)$ is not available, we use solely the measured
$F(H\gamma)/F(H\beta)$ ratio to obtain the reddening
correction using

\begin{equation}
E(B - V) =  4.71 \times \left[ {Log\left( {\frac{{F(H\gamma
)}}{{F(H\beta )}}} \right) - Log(0.47)} \right]
\end{equation}

The resulting color excess is used to correct each spectrum with
the IRAF task $deredden$. E(B-V) is easily converted to $cH\beta$
using the relation $E(B-V)\approx 0.677\times cH\beta$ (Vogel et
al. 1993).

In Table 2 we compare the reddening E(B-V) measured in this work
with that obtained by other authors. Furthermore, we have compared
our derived color excesses with other measurements taken from the
literature in Figure 2. The top panel shows this comparison, while
the bottom panel shows the deviations as a function of E(B-V). The
median difference is 0.06 with a standard deviation of 0.18.

We note some differences between the values obtained for some of
the objects. The spectroscopic results taken from the literature
are inhomogeneous in their instrumental accuracy, observational
techniques, and the S/N ratios of individual observations.
However, it is unlikely that instrumental differences dominate
over variations in measured lines intensities due to observing
emission from different locations within each galaxy or star
formation region, different sized apertures, or slightly different
position angles used by the authors.

\setcounter{table}{1}
\begin{table}
\begin{minipage}{220mm}
\caption{E(B-V): Comparison with Literature.}
\begin{tabular}{lccc}
\hline\hline

&  Observed&  Literature& Reference\\

\hline

Fairall44&     0.47& 0.60& 5\\

III Zw 107&    0.53& 0.47& 9\\

Mrk309&        0.73& 0.71, 0.63, 0.70& 9, 2, 12\\

Mrk475&        0.21& 0.10& 7\\

Mrk712&        0.03& 0.18& 3\\

Mrk1271&       0.15& 0.10& 8\\

NGC450&        0.55&  -& -\\

NGC4385&       0.64& 0.60, 0.52& 2, 14\\

NGC4861&       0.43& 0.20, (0.49, 0.62, 0.77)& 2,(1)\\

NGC5430&       0.67& 1.14& 4\\

NGC5471&       0.21& 0.15& 13\\

NGC6764&       0.82,0.80&  0.61,(0.87)& 10, (6)\\

NGC7714&       0.40&  (0.39,041), 0.43& (2), 11\\

UM48&          0.57& 0.49& 15\\

\hline
\end{tabular}

\end{minipage}
(1) Barth et al. 1994, (2) Calzetti 1997, (3) Contini et al. 1995,
(4) Contini et al. 1997, (5) Durret 1990, (6) Eckart et al. 1996,
(7) Izotov et al. 1994, (8) Izotov \& Thuan 1998,  (9) Gil de Paz
et al. 2000, (10) Keel 1982, (11) Mattila \& Meikle 2001, (12)
Osterbrock \& Cohen 1982, (13) Rosa \& Bevenuti 1994, (14) Salzer
et al. 1989 ,(15) Vogel et al. 1993

\end{table}

\setcounter{figure}{1}
\begin{figure*}
\includegraphics[width=150mm]{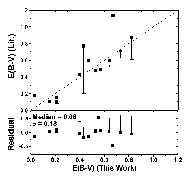} \caption{Top:
Comparison of the values of E(B-V) adopted in this work and those
obtained by other authors. The dashed line shows a relation of
unity. Bottom: The difference between the values adopted in this
work and those obtained  by other authors.} \label{fig5}
\end{figure*}

After correcting the spectra for the total interstellar extinction
using the Balmer line ratios, some objects show
$F(H\gamma)/F(H\beta)$ ratios lower than the theoretically
predicted values. This effect is more significant for objects
where the H$\beta$ equivalent width is less than $70$\AA~ (McCall,
Rybski \& Shields 1985, hereafter MRS85). The consequence of such
underlying absorption is an overestimate of the reddening
correction, which is then reflected in all the observed emission
lines. To correct for this effect, we use the method proposed by
MRS85 to compute the equivalent width in absorption, $EW_{abs}$,
given by the expression,

\begin{equation}
\left(\frac{{F(H\gamma )}}{{F(H\beta )}}\right)_{dered} =
\frac{{I(H\gamma )}}{{I(H\beta )}}\frac{{\left(1 +  {EW_{abs}
/EW(H\beta) } \right)^{1 - \varepsilon } }}{{\left(1 +  {EW_{abs}
/EW(H\gamma)} \right)}}.
\end{equation}

The parameter $\varepsilon$ is defined by

\begin{equation}
\varepsilon  = \frac{{\left( {\frac{{A(H\gamma )}}{{A_V }} -
\frac{{A(H\beta )}}{{A_V }}} \right)}}{{\left( {\frac{{A(H\alpha
)}}{{A_V }} - \frac{{A(H\beta )}}{{A_V }}} \right)}}
\end{equation}

\noindent and represents the extinction reddening law (CCM89 and
MRS85).

In the limit $EW_{abs} = 0$, $(F(H\gamma)/F(H\beta))_{dered}$ reaches
 the theoretical ratio ($I(H\gamma)/I(H\beta)=0.47$). In the opposite
 limit, as $EW_{abs}$ approaches unity, the ratio
 $(F(H\gamma)/F(H\beta))_{dered}$ is significantly altered from the
 theoretical expectation. The corrected equivalent widths of the
 Balmer lines are obtained by adding $EW_{abs}$ to the observed
 equivalent widths.

The correction in the extinction due to underlying absorption
lines, E(B-V)$_{abs}$, is given by the expression

\begin{equation}
 E(B - V)= - 2.33 \times Log\left( {\frac{{1 + \left( {EW_{abs} /EW(H\alpha) }
 \right)}}{{1 + \left( {EW_{abs} /EW(H\beta)  } \right)}}} \right)
\end{equation}

The value of $EW_{abs}$ for each object is listed in Table 3,
along with the equivalent widths of the emission lines $EW(H\alpha
)$, $EW(H\beta )$ and $EW(H\gamma )$. Reddening and underlying
absorption corrected intensities are also given in the same Table.

\setcounter{table}{2}
\begin{table*}
\centering
\begin{minipage}{220mm}
\caption{Corrected Emission Line Intensities Relative to $H\beta
=1000$}
\begin{tabular}{lrrrrrr}

\hline\hline

& Fairall44& IIIZW107& MRK309& MRK475& Mrk712& Mrk1271\\

\hline

[OII]$\lambda$3727&           -&        2213$\pm$12&      703$\pm$98&     1093$\pm$4&          -&            -\\

[NeIII]$\lambda$3869&         -&          259$\pm$5&               -& 362.9$\pm$2.7&          -&             -\\

H$\gamma$&            475$\pm$2&          470$\pm$5&      466$\pm$46&      470$\pm$3&  470$\pm$3&      470$\pm$8\\

[OIII]$\lambda$4363&   26$\pm$9&           13$\pm$4&               -&       85$\pm$3&   14$\pm$2&       61$\pm$12\\

HeI$\lambda$4471&     53$\pm$9&           49$\pm$5&       59$\pm$38&       42$\pm$2&   67$\pm$2&       61$\pm$15\\

[FeIII]$\lambda$4656&   29$\pm$5&          13$\pm$7&      131$\pm$29&        5$\pm$3&   14$\pm$3&        26$\pm$7\\

HeII4686&               19$\pm$5&           3$\pm$9&               -&       17$\pm$4&          -&       35$\pm$11\\

[ArIV]$\lambda$4711&           -&         14$\pm$10&               -&        8$\pm$3&    8$\pm$2&               -\\

[ArIV]$\lambda$4740&           -&                 -&               -&        7$\pm$3&          -&             -\\

HeI$\lambda$4922&              -&          12$\pm$8&               -&       11$\pm$3&   19$\pm$2&            -\\

[OIII]$\lambda$4959&   710$\pm$5&        1215$\pm$7&      104$\pm$25&     1781$\pm$5& 1613$\pm$4&     1115$\pm$12\\

[OIII]$\lambda$5007&  2061$\pm$9&       3632$\pm$18&      381$\pm$33&    5326$\pm$15& 4881$\pm$11&     3096$\pm$29\\

[NI]$\lambda$5199&                -&           31$\pm$7&          -&                  4$\pm$3&       12$\pm$3&                    -\\

[FeIII]$\lambda$5271&             -&                      -&          -&                 7$\pm$3&       10$\pm$3&                    -\\

[ClIII]$\lambda$5518&             -&           10$\pm$2&          -&                     8$\pm$2&        8$\pm$5&                    -\\

[ClIII]$\lambda$5538&             -&            3$\pm$1&          -&                    4$\pm$2&        6$\pm$5&                   -\\

[NII]$\lambda$5754&               -&            9$\pm$2&       38$\pm$25&        5$\pm$3&        7$\pm$3&              -\\

HeI$\lambda$5876&    126$\pm$4&          129$\pm$2&       24$\pm$17&       99$\pm$1&      228$\pm$3&      119$\pm$9\\

[OI]$\lambda$6300&     48$\pm$6&           53$\pm$2&       26$\pm$22&       24$\pm$1&       36$\pm$3&      65$\pm$11\\

[SIII]$\lambda$6313&     8$\pm$4&           11$\pm$2&                 -&       24$\pm$1&       24$\pm$3&       15$\pm$10\\

[OI]$\lambda$6364&     18$\pm$4&           18$\pm$3&                -&               9$\pm$1&       12$\pm$3&       72$\pm$11\\

[NII]$\lambda$6548&               -&           98$\pm$2&               -&       33$\pm$2&       268$\pm$2&       84$\pm$11\\

H$\alpha$&                        -&        2855$\pm$14&               -&     2870$\pm$8&                 -&     2899$\pm$25\\

[NII]$\lambda$6584&               -&          280$\pm$2&               -&       66$\pm$1&                 -&      221$\pm$9\\

\hline

EW(H$\gamma$)(\AA)&          32.17$\pm$0.23&         26.65$\pm$0.35&       3.00$\pm$0.18&      73.08$\pm$3.35&      45.98$\pm$0.21&      15.42$\pm$0.20\\

EW(H$\beta)(\AA)$&           67.64$\pm$0.26&         65.48$\pm$0.46&      10.18$\pm$0.79&     140.50$\pm$0.37&     139.10$\pm$0.29&      33.11$\pm$0.28\\

EW(H$\alpha)(\AA)$&                       -&       329.60 $\pm$3.70&          -&            936.20$\pm$10.98&          -&         109.20$\pm$1.16\\

EW$_{abs}(\AA)$&                         -&                   -0.98&                      -&                      -1.18&            -&               -0.69\\

F(H$\beta$)(10$^{-14}$ erg s$^{-1}$ cm$^{-2}$)&             3.58$\pm$0.01&         12.77$\pm$0.06&   12.8$\pm$1.04&   7.52$\pm$0.03&   2.45$\pm$0.02&   0.72$\pm$0.01\\

E(B-V)&                         0.47&                  0.53&                  0.73&                  0.21&                   0.03&                  0.15 \\

E(B-V)$_{abs}$&                    -&                -0.012&                     -&                -0.007&                      -&                 -0.015  \\

\hline\hline

&                        &                  &            Broad Emission Lines&                     &           &               \\

\hline

NIII$\lambda$4640&                  -&            7$\pm$5&      167$\pm$33&          -&             82$\pm$10&                    -\\

HeII$\lambda$4886&      59$\pm$9&          61$\pm$26&      115$\pm$20&       69$\pm$16&       98$\pm$7&      103$\pm$41\\

CIII$\lambda$5696&                 -&                      -&       15$\pm$4&                  -&                   -&          -\\

CIV$\lambda$5808&                  -&                      -&          -&                 36$\pm$7&                  -&                 -\\

EW$\lambda$4640(\AA)&                    -&          0.37$\pm$0.38&       1.36$\pm$0.21&                 -&       7.23$\pm$0.98&                   -\\

EW$\lambda$4686(\AA)&        3.47$\pm$0.56&          3.60$\pm$2.86&       0.97$\pm$0.20&       9.51$\pm$2.35&       8.67$\pm$0.74&       3.06$\pm$1.21\\

EW$\lambda$5696(\AA)&                    -&                      -&          0.26$\pm$0.05&                -&                 -&                   -\\

EW$\lambda$5808(\AA)&                    -&                      -&                    -&      8.02$\pm$1.46&                 -&                   -\\

\hline

\end{tabular}
\end{minipage}
\end{table*}

\setcounter{table}{2}
\begin{table*}
\centering
\begin{minipage}{220mm}
\caption{Continued}
\begin{tabular}{lrrrrrr}

\hline\hline

& NGC450& NGC4385& NGC4861& NGC5430& NGC5471& NGC6764\\

\hline

[OII]$\lambda$3727&     1615$\pm$3&               -&      626$\pm$2&     1491$\pm$28&      919$\pm$2&     1933$\pm$29\\

[NeIII]$\lambda$3869&      240$\pm$2&              -&      408$\pm$2&           -&      399$\pm$2&       68$\pm$18\\

H$\gamma$&      481$\pm$2&      471$\pm$3&      470$\pm$2&      470$\pm$23&      470$\pm$2&      470$\pm$12\\

[OIII]$\lambda$4363&       27$\pm$2&       13$\pm$3&       89$\pm$2&          -&       66$\pm$2&       21$\pm$8\\

HeI$\lambda$4471&       48$\pm$2&       30$\pm$6&       37$\pm$2&       46$\pm$9&       39$\pm$2&       47$\pm$13\\

[FeIII]$\lambda$4656&        9$\pm$3&       35$\pm$6&        8$\pm$3&       49$\pm$23&       26$\pm$4&       69$\pm$22\\

HeII$\lambda$4686&          -&       14$\pm$4&        8$\pm$2&       32$\pm$10&       25$\pm$1&       54$\pm$15\\

[ArIV]$\lambda$4711&        5$\pm$2&          -&       11$\pm$2&          -&       14$\pm$1&        -\\

[ArIV]$\lambda$4740&          -&          -&        7$\pm$2&          -&          -&          -\\

HeI$\lambda$4922&       13$\pm$2&         -&        5$\pm$2&          -&        9$\pm$1&          -\\

[OIII]$\lambda$4959&     1368$\pm$3&      231$\pm$3&     1817$\pm$4&      129$\pm$10&     1850$\pm$30&      168$\pm$6\\

[OIII]$\lambda$5007&      3864$\pm$7&      678$\pm$4&     5521$\pm$11&      375$\pm$12&     5583$\pm$90&      484$\pm$8\\

[NI]$\lambda$5199&        9$\pm$2&       19$\pm$3&        3$\pm$3&       12$\pm$12&          -&       88$\pm$14\\

[FeIII]$\lambda$5271&        6$\pm$3&          -&        -&          -&        2$\pm$2&          -\\

[ClIII]$\lambda$5518&        6$\pm$1&          -&        4$\pm$2&        1$\pm$1&        5$\pm$1&        7$\pm$5\\

[ClIII]$\lambda$5538&        5$\pm$1&          -&        3$\pm$2&        2$\pm$2&        6$\pm$2&          -\\

[NII]$\lambda$5754&        2$\pm$1&        10$\pm$3&          -&       12$\pm$4&        3$\pm$2&       29$\pm$7\\

HeI$\lambda$5876&      131$\pm$1&      127$\pm$3&       92$\pm$1&      119$\pm$5&      111$\pm$1&      123$\pm$6\\

[OI]$\lambda$6300&       25$\pm$1&       25$\pm$4&       16$\pm$2&          -&       22$\pm$1&      159$\pm$7\\

[SIII]$\lambda$6313&       15$\pm$1&        10$\pm$4&       17$\pm$2&        6.7$\pm$3.3&       18.6$\pm$0.7&          -\\

[OI]$\lambda$6364&        9$\pm$1&        6$\pm$5&        5$\pm$1&        4$\pm$3&        8$\pm$1&       67$\pm$8\\

[NII]$\lambda$6548&       71$\pm$1&      503$\pm$4&       23$\pm$2&      441$\pm$11&       46$\pm$1&      726$\pm$12\\

H$\alpha$&     2872$\pm$5&     2884$\pm$8&     2855$\pm$22&     2837$\pm$49&     2866$\pm$5&     2806$\pm$35\\

[NII]$\lambda$6584&      167$\pm$1&     1489$\pm$5&       47$\pm$2&     1207$\pm$22&       93$\pm$1&     1914$\pm$24\\

\hline

EW(H$\gamma)(\AA)$&   93.61$\pm$0.31&      13.85$\pm$0.09&      48.25$\pm$0.32&      13.21$\pm$0.45&      51.68$\pm$0.18&   6.58$\pm$0.14\\

EW(H$\beta$)(\AA)&    320.80$\pm$0.59&     42.78$\pm$0.11& 156.80$\pm$0.29&     40.85$\pm$0.41&  181.20$\pm$0.29& 22.46$\pm$0.27\\

EW(H$\alpha)(\AA)$& 2928.00$\pm$10.59&     199.10$\pm$1.66&     938.00$\pm$5.61&     264.60$\pm$1.25&    1028.00$\pm$6.35&     136.40$\pm$0.90\\

EW$_{abs}(\AA)$&                 -0.27&                   -2.18&                0.00&                    -0.05&                  -0.40&                  -0.72\\

F(H$\beta$)(10$^{-14}$ erg s$^{-1}$ cm$^{-2}$)&  30.78$\pm$0.06&      48.63$\pm$0.15&     103.60$\pm$0.20&      37.17$\pm$0.63&       7.73$\pm$0.01&      59.81$\pm$0.83\\

E(B-V)&               0.55&                    0.64&                    0.43&                    0.67&                    0.21&                    0.82\\

E(B-V)$_{abs}$&          -0.001&               -0.039&                   0.000&                 -0.001&                  -0.002&                  -0.027\\

\hline\hline

&                        &                  &            Broad Emission Lines&                     &           &               \\

\hline

NIII$\lambda$4640&     21$\pm$16&       95$\pm$11&          -&      127$\pm$26&          -&      161$\pm$33\\

HeII$\lambda$4886&    31$\pm$14&      103$\pm$8&       55$\pm$13&      105$\pm$18&       22$\pm$4&       63$\pm$21\\

CIII$\lambda$5696&                -&       31$\pm$31&        3$\pm$2&          -&          -&       10$\pm$5\\

CIV$\lambda$5808&      23$\pm$3&       62$\pm$34&       11$\pm$6&       41$\pm$10&          -&       74$\pm$27\\

EW$\lambda$4640(\AA)&       5.14$\pm$3.80&       3.53$\pm$0.39&          -&       5.50$\pm$0.94&          -&       2.94$\pm$0.61\\

EW$\lambda$4686(\AA)&       7.61$\pm$3.80&       3.91$\pm$0.26&       7.08$\pm$1.75&       3.61$\pm$0.78&       3.14$\pm$0.69&       1.15$\pm$0.57\\

EW$\lambda$5696(\AA)&                   -&       2.16$\pm$2.12&       0.81$\pm$0.47&          -&          -&       0.30$\pm$0.18\\

EW$\lambda$5808(\AA)&      10.58$\pm$1.60&       4.55$\pm$2.41&       3.00$\pm$1.49&       3.16$\pm$1.77&          -&       3.90$\pm$0.76\\

\hline
\end{tabular}
\end{minipage}
\end{table*}

\setcounter{table}{2}
\begin{table*}
\centering
\begin{minipage}{220mm}
\caption{Continued}
\begin{tabular}{lrrrr}

\hline\hline

& NGC6764P.A.90& NGC7714A& NGC7714B& UM48\\

\hline

[OII]$\lambda$3727&     1703$\pm$31&     1970$\pm$8&     2107$\pm$10&     2152$\pm$29 \\

[Ne III]$\lambda$3869&      130$\pm$23&       95$\pm$3&      280$\pm$4&      111$\pm$15 \\

H$\gamma$&            471$\pm$15&      472.4$\pm$3.7&      469.9$\pm$4.4&      472$\pm$16 \\

[O III]$\lambda$4363&       19$\pm$8&       13.6$\pm$4.2&       39.3$\pm$4&       12$\pm$6 \\

HeI$\lambda$4471&       59.7$\pm$16.5&       39.6$\pm$3.5&       46$\pm$4&       54$\pm$18 \\

[FeIII]$\lambda$4656&       85$\pm$26&       30$\pm$5&       28$\pm$9&       35$\pm$23 \\

HeII$\lambda$4686&       58$\pm$23&        6$\pm$1&        4$\pm$4&       12$\pm$16 \\

[ArIV]$\lambda$4711&          -&          -&       22$\pm$7&          - \\

[ArIV]4740&          -&          -&          -&          -\\

HeI$\lambda$4922&          -&          -&       12$\pm$4&          -\\

[OIII]$\lambda$4959&      171$\pm$8&      496$\pm$40&     1180$\pm$6&      518$\pm$14\\

[OIII]$\lambda$5007&      484$\pm$11&     1550$\pm$140&     3529$\pm$14&     1563$\pm$20\\

[NI]$\lambda$5199&       77$\pm$15&       18$\pm$4&          -&          -\\

[FeIII]$\lambda$5271&          -&       27$\pm$4&       23$\pm$4&          -\\

[ClIII]$\lambda$5518&       10.7$\pm$9&        7$\pm$4&          -&          -\\

[ClIII]$\lambda$5538&        4.2$\pm$6&        7$\pm$4&          -&          -\\

[NII]$\lambda$5754&       20$\pm$7&       11$\pm$3&       11$\pm$10&          -\\

HeI$\lambda$5876&        99$\pm$7&      125$\pm$3&      117$\pm$3&      125$\pm$5\\

[OI]$\lambda$6300&      156$\pm$8&       44$\pm$4&       46$\pm$4&       34$\pm$10\\

[SIII]$\lambda$6313&          -&        8$\pm$4&       15$\pm$4&       15$\pm$11\\

[OI]$\lambda$6364&       58$\pm$8&       14$\pm$6&       16$\pm$4&       13$\pm$7\\

[NII]$\lambda$6548&      708$\pm$13&      362$\pm$4&      131$\pm$4&      197$\pm$5\\

H$\alpha$&       2811$\pm$40&     2823$\pm$123&     2872.2$\pm$12&     2869$\pm$31\\

[NII]$\lambda$6584&     1912$\pm$28&      923$\pm$5&      334$\pm$4&      573$\pm$8\\

\hline

EW(H$\gamma$)(\AA)&       6.39$\pm$0.17&      12.13$\pm$0.08&      68.88$\pm$0.57&       9.95$\pm$0.31\\

EW(H$\beta$)(\AA)&      22.04$\pm$0.30&      32.99$\pm$0.10&     191.30$\pm$0.72&      29.25$\pm$0.31\\

EW(H$\alpha$)(\AA)&     137.70$\pm$0.97&      98.48$\pm$0.63&      12.00$\pm$0.01&     156.30$\pm$1.09\\

EW$_{abs}$(\AA)&       -0.94&       -1.44&       -0.18&       -1.29\\

F(H$\beta$)(10$^{-14}$ erg s$^{-1}$ cm$^{-2}$)&             48.98$\pm$0.80&      86.15$\pm$0.32&       7.79$\pm$0.03&       9.32$\pm$0.12\\

E(B-V)&                   0.80&                   0.52&                   0.40&                   0.57\\

E(B-V)$_{abs}$&                 -0.036&                 -0.029&                 -0.014&                 -0.035\\

\hline \hline

&                                     Broad Emission Lines&                     &           &               \\

\hline

NIII$\lambda$4640&      161$\pm$39&       41$\pm$9&          -&          -\\

HeII$\lambda$4886&       64$\pm$23&       22$\pm$10&       54$\pm$14&       56$\pm$33\\

CIII$\lambda$5696&       10.$\pm$7&        5.0$\pm$0.6&          -&          -\\

CIV$\lambda$5808&       81$\pm$29&       34$\pm$14&          -&       22$\pm$12\\

EW$\lambda$4640(\AA)&                   2.94$\pm$0.71&       1.41$\pm$0.25&          -&           -\\

EW$\lambda$4686(\AA)&                   1.20$\pm$0.54&       0.60$\pm$0.27&      10.34$\pm$2.92&       1.50$\pm$0.84\\

EW$\lambda$5696(\AA)&                   0.23$\pm$0.17&       0.22$\pm$0.03&          -&          -\\

EW$\lambda$5808(\AA)&                   3.65$\pm$0.69&       1.77$\pm$0.08&          -&       1.52$\pm$0.71\\

\hline
\end{tabular}
\end{minipage}
\end{table*}

\section{Hard Ionization Sources}
\label{section:ionization}

The contamination of a starburst by nearby objects such as SN IIe
remnants and AGN could occur, producing Wolf-Rayet-like features
in the spectrum (Masegosa et al. 1991). Embedded or superimposed
supernova remnants (SNRs) can strongly affect the emission-line
properties in a predominantly photoionized HII region
($\it{e.g.}$, Peimbert, Sarmiento, \& Fierro 1991). It is crucial
to distinguish between hot stars and other ionizing photon
sources. From an observational viewpoint, the presence of shocks
in an ionized nebula can be detected by the enhancement of low
excitation lines, in particular [OI], [SII] and [NII] (Masegosa et
al. 1991).

We used Veilleux {\&} Osterbrock (1987) diagnostic diagrams to
verify the ionization sources in our galaxy sample. Since
[SII]$\lambda\lambda6717,6731$ are not available in our spectra,
we use diagnostic diagrams plotting [OI]$\lambda6300/$H$\alpha$
and [NII]$\lambda6584/$H$\alpha$ line ratios against the
[OIII]$\lambda5007/$H$\beta$ ratios (Figure 3 and 4).

The theoretical starburst limits shown in Figure 3 and 4 can be
parameterized as:

\begin{equation}
\frac{{log([OIII]\lambda5007) }}{{H\beta }} = \frac{{0.61
}}{{log([NII]\lambda6584/H\alpha) }-0.47}+1.19
\end{equation}

\begin{equation}
\frac{{log([OIII]\lambda5007 }}{{H\beta }} = \frac{{0.73
}}{{log([OI]\lambda6300/H\alpha) }+0.59}+1.33
\end{equation}

\noindent given by Kewley et al. (2001). The error range of their modeling
in both planes is $\pm$ 0.1 dex.

All objects in our sample are located in the starburst and HII region
loci in the plane of [OI]$\lambda6300/$H$\alpha$ and
[NII]$\lambda6584/$H$\alpha$ versus [OIII]$\lambda5007/$H$\beta$
(Veilleux \& Osterbrock 1987), suggesting that the main sources of
ionizing photons are stellar, and not AGN.

\setcounter{figure}{2}
\begin{figure*}
\includegraphics[width=150mm]{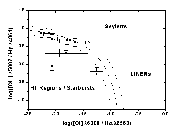} \caption{The diagnostic
diagram of [OIII]$\lambda5007/$H$\beta$ plotted against
[OI]$\lambda6300/$H$\alpha$.The solid line shows the border
between starburst and AGNs ionization mechanisms. The dashed lines
show the model uncertainty of 0.1 dex.} \label{fig2}
\end{figure*}

\setcounter{figure}{3}
\begin{figure*}
\includegraphics[width=150mm]{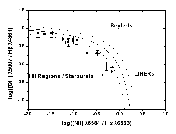} \caption{The diagnostic
diagram of [OIII]$\lambda5007/$H$\beta$ plotted against
$[NII]\lambda6584/$H$\alpha$.The solid line shows the border
between starburst and AGNs ionization mechanisms. The dashed lines
show the model uncertainty of 0.1 dex.} \label{fig3}
\end{figure*}

\section{Gas Chemical Abundances}
\label{section:abundance}

The oxygen abundance O/H is a crucial parameter for our study
since one of the main purposes of this paper is to study the
massive star populations in galaxies with different metallicities.
In order to obtain the chemical abundances of the galaxies in our
sample, three subclasses were defined depending on whether or not
the [OIII]$\lambda4363$, [OII]$\lambda3727$ and [OIII]$\lambda
\lambda4959,5007$ features are present (see Kobulnicky et al.
1999):

\noindent Class 1. [OIII]$\lambda4363$, [OII]$\lambda3727$ and
[OIII]$\lambda \lambda4959,5007$ are detected. When
[OIII]$\lambda4363$ is detected, the temperature is obtained from the
[OIII]$\lambda\lambda5007,4959/\lambda4363$ line ratio, while the
density is measured from the [SII]$\lambda6731$/[SII]$\lambda6717$
line ratio (McCall 1984). Since our spectra do not include the [SII]
$\lambda\lambda6717,6731$ lines, we are unable to measure the electron
density, and we assume a fiducial value of 150 cm$^{- 3}$ for $n_e$.

The accuracy of this empirical method depends strongly on an
accurate estimate of the electron temperature of the gas
(Steigman, Viegas {\&} Gruenwald 1997). The uncertainty in the
electron temperature of the gas is directly related to the
measurement of the [OIII]$\lambda4363$ line intensity. The oxygen
abundance can be obtained by the empirical method first proposed
by Peimbert {\&} Costero (1969), which requires knowledge of the
gas temperature and density.

\noindent Class 2. [OIII]$\lambda4363$ is not detected, while
[OII]$\lambda3727$ and [OIII]$\lambda \lambda4959,5007$ are
detected. For those galaxies where the [OIII]$\lambda4363$ auroral
line is not detected, the oxygen abundance is obtained from the
empirical calibration suggested by McGaugh (1991) using the known
R$_{23}$ ratio,
([OII]$\lambda3727$+[OIII]$\lambda\lambda5007,4959)/H\beta$.

The relation between oxygen abundance and R$_{23}$ is double
valued, but this degeneracy can be broken. We use the relation
between [NII]$\lambda6584$/[OII]$\lambda3727$ and
[NII]$\lambda6584$/H$\alpha$ ratios to break the O/H vs. R$_{23}$
degeneracy (Contini et al. 2002).

Kobulnicky et al. (1999) provide a polynomial fit to both
metal-poor (lower branch),
\begin{equation}
12+log(O/H)_{l} = 12 - 4.944 + 0.767x + 0.0602x^{2} - y(0.29 +
0.332x - 0.331x^{2}),
\end{equation}
and metal-rich (upper branch) regimes,
\[
12+log(O/H)_{u} = 12 - 2.939 - 0.2x - 0.237x^{2} - 0.305x^{3} -
0.0283x^{4}
\]

\begin{equation}
 - y(0.0047 - 0.0221x - 0.102x^{2} - 0.0817x^{3} -
0.00717x^{4}),
\end{equation}

\noindent where
\begin{equation}
x\equiv log R_{23}\equiv
log([OII]\lambda3727+[OIII]\lambda\lambda5007,4959)/H\beta)
\end{equation}
\noindent and
\begin{equation}
y\equiv log O_{32}\equiv
log([OIII]\lambda\lambda5007,4959)/[OII]\lambda3727).
\end{equation}

The $R_{23}$ calibration has an estimated uncertainty of $\pm0.10$
in log(O/H) (Kobulnicky et al. 1999).

\noindent Class 3. [OII]$\lambda3727$ is not detected, but
[OIII]$\lambda4363$ and [OIII]$\lambda \lambda4959,5007$ are
detected (NTT data). It is not possible to break the degeneracy
using the [NII]$\lambda6584$/[OII]$\lambda3727$ and
[NII]$\lambda6584$/H$\alpha$ ratios.

In this case, we can derive the electron temperature but not the
O/H abundance. We use the temperature obtained from standard
nebular analysis to select the appropriate metallicity calibration
with which we derive the oxygen abundance.  After determining the metallicity
regime for each galaxy, we compute the O/H ratio using the upper
branch given by Edmunds {\&} Pagel (1984),
\begin{equation}
log(O/H)_{u} \simeq -0.69 log R_3 -3.24,
\end{equation}
\[ (0.6 \leq log R_3 \leq 1.0)
\]

\[ R_3 \equiv
\frac{{F_{[OIII]\lambda \lambda\ 5007,4959}}}{{F_{H\beta }}}
\]

\noindent for galaxies in the high metallicity regime, and the lower branch

\begin{equation}
log(O/H)_{l} \simeq 1.67 log R_3 + 6.43,
\end{equation}
\[ (0.4 \leq log R_3 \leq 1.1)
\]

\noindent for low metallicity objects. The $R_{3}$ calibration has an
estimated uncertainty of $\pm0.20$ in $log(O/H)$ (Edmunds {\&}
Pagel 1984).

Two additional empirical tools were used to obtain the oxygen
abundance: the P-R$_{23}$ calibration, with an estimated
uncertainty of 0.10 dex (Pilyugin 2000, 2001a and 2001b), and the
[OIII]/[NII] calibration given by Pettini and Pagel (2004), with
an uncertainty of 0.25 dex. In particular, for NTT data we use the
calibration extrapolated to [OII]$\lambda3727=0$, where the
uncertainty increases to 0.30 dex (Pilyugin 2001a).

To compare the derived oxygen abundances, we use as many of the
methods described above as possible (based on available spectral
features) to derive O/H values for all of the galaxies in our
sample. Figure 5 shows the comparison between results obtained
applying the T$_e$-Method and the other methods used in this work.
We see that in the absence of [OIII]$\lambda$4363, use of the
electron temperature T$_e$, the R$_{23}$ method (for Class 2
objects), and the [OIII]/[NII] calibration all provide good
determinations of gas abundance. The R$_{3}$ technique was used
only for MRK 712. We find that the R$_{23}$ and R$_{3}$
calibrations of Pilyugin do not perform correctly for either high
or very low gas abundances, and are therefore not used in this
work.

Furthermore, we have compared our derived gas abundances with
those in the literature in Figure 6. The top panel shows this
comparison, while the bottom panel shows the deviations as a
function of 12+log(O/H).The median difference is 0.02 dex with a
standard deviation of 0.17 dex.

The above comparisons, both internal and external, demonstrate
that our gas abundance determinations are reliable and robust. The
gas abundances must be measured correctly to accurately constrain
the massive stellar population and its relation to gas
metallicity. The abundances obtained using the different methods
and the final adopted value for the oxygen abundance of each
galaxy are listed in Table 4.

\setcounter{figure}{4}
\begin{figure*}
\includegraphics[width=150mm,height=180mm]{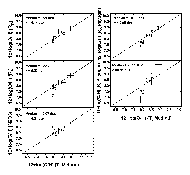} \caption{Comparison
between the values of 12 + log(O/H) obtained with the T$_e$ method
and other methods used in this work. The median offset and
$\sigma$ for each relation are noted in each panel. The R$_{23}$ and
R$_{3}$ calibrations of Pilyugin are shown separately on the left.}
\label{fig4}
\end{figure*}

\setcounter{figure}{5}
\begin{figure*}
\includegraphics[width=150mm,height=180mm]{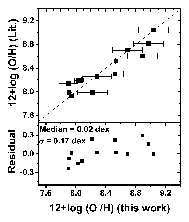} \caption{Top:
Comparison of the values of 12 + log(O/H) adopted in this work and
those obtained by other authors. The dashed line shows a relation
of unity. Bottom: The difference between the values obtained  by
other authors and those adopted in this work.} \label{fig5}
\end{figure*}

\setcounter{table}{3}
\begin{table}
\begin{minipage}{220mm}
\caption{Oxygen abundance obtained using different methods.}
\begin{tabular}{lccccccccccc}

\hline\hline

& & & & & Method& & & & \\

\hline

& R Branch& T$_e$&  R$_{23}$& N2O3& R$_3$& R$_{23}$ Pilyugin&
R$_3$ Pilyugin& Adopted& $\Delta$ Max& Literature\\

\hline

Fairall44&   upper&                  -&      -&      -&   8.45&      -&   8.26&   8.45$\pm$0.20& 0.19&       -\\

III Zw107&   upper&      8.52$\pm$0.12&   8.57&   8.23&   8.29&   8.31&   8.39&   8.52$\pm$0.12& 0.34& 8.30(d)\\

Mrk309&      upper&                  -&   9.04&      -&   8.98&   9.40&  10.07&   9.04$\pm$0.20& 1.09& 9.03(h)\\

Mrk475&      lower&      7.94$\pm$0.01&   8.00&   7.97&   7.95&   7.81&   7.58&   7.94$\pm$0.01& 0.39& 7.93(e)\\

Mrk712&      upper&                  -&      -&      -&   8.28&      -&   8.37&   8.28$\pm$0.20& 0.09& 8.26(a)\\

Mrk1271&     lower&                  -&      -&   8.22&   7.57&      -&   8.49&   8.22$\pm$0.20& 0.92& 7.99(e)\\

NGC450&      upper&      8.23$\pm$0.03&   8.61&   8.15&   8.26&   8.33&   8.33&   8.23$\pm$0.03& 0.18&    -\\

NGC4385&     upper&                  -&      -&   8.69&   8.79&      -&   9.61&   8.69$\pm$0.20& 0.92& 8.70(i)\\

NGC4861&     lower&      7.91$\pm$0.02&   7.89&   7.92&   7.98&   7.79&   7.61&   7.91$\pm$0.02& 0.37& 7.99(e)\\

NGC5430&     upper&                  -&   8.97&   8.75&   8.97&   9.08&  10.04&   8.97$\pm$0.20& 1.29& 8.81(b)\\

NGC5471&     lower&      8.08$\pm$0.01&   7.99&   8.01&   7.99&   7.82&   7.61&   8.08$\pm$0.01& 0.38& 8.20(c)\\

NGC6764&     upper&      8.89$\pm$0.20&   8.90&   8.78&   8.89&   8.92&   9.85&   8.89$\pm$0.20& 1.07& 8.60(b)\\

NGC6764& upper&          8.89$\pm$0.20&   8.93&   8.78&   8.89&   8.98&   9.85&   8.89$\pm$0.20& 1.07&    -\\

NGC7714A&    lower&      7.90$\pm$0.13&   7.73&   8.51&   7.05&   7.38&   6.80&   7.90$\pm$0.13& 1.71& 8.14(f)\\

NGC7714B&    lower&      8.03$\pm$0.05&   8.00&   8.26&   7.65&   7.70&   7.33&   8.03$\pm$0.05& 0.93& 8.19(f)\\

UM48&        upper&      8.54$\pm$0.01&   8.76&   8.44&   8.54&   8.62&   9.01&   8.54$\pm$0.01& 0.57& 8.52(g)\\

\hline
\end{tabular}
\end{minipage}
(a) Contini et al. (1995), (b) Contini et al. (1997), (c) Evans
(1986), (d) Gil de Paz et al. (2000), (e) GIT2000, (f)
Gonzalez-Delgado et al. (1995), (g) Masegosa et al. (1991), (h)
SGTI 2000, (i) Sugai et al. (1992).
\end{table}

\section{Massive Star Populations: WR and O Star Numbers}
\label{section:hotstars}

The number of massive stars present in a region of a Wolf-Rayet
galaxy can be derived using the standard method developed by Conti
(1991). A representative number of O and WR subtype stars can be
obtained directly from the optical spectra (Vacca {\&} Conti 1992,
hereafter VC92).

The absolute number of WR stars in the sampled galaxies was
estimated using the blue bump ($\lambda $4686) and red bump
($\lambda $5808) luminosities. The blue bump is a blend of
NV$\lambda4605$, NIII$\lambda\lambda4634,4640$, CIII$\lambda4650$,
CIV$\lambda4658$ and HeII$\lambda4686$ broad WR lines (GIT00).
Superposed on the blue bump may be [FeIII]$\lambda4658$,
HeII$\lambda4686$, and [ArIV]$\lambda\lambda 4711,4740$, which are
narrow nebular lines. The dominant contribution to the blue bump
is from the broad HeII$\lambda$4686 line arising in WNL stars,
while the red bump is mainly due to broad CIV $\lambda $5808 from
WC stars (Schaerer et al. 1999, GIT00). However, some contribution
from early WN stars (WNE) might be present in the blue bump
(SV98). The detectability of the red bump is low compared to the
blue bump. Figure 7 shows the spectral region of those galaxies in
our sample containing either the blue or red bump.

\setcounter{figure}{6}
\begin{figure*}
\includegraphics[width=180mm,height=220mm]{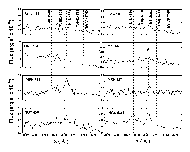} \caption{Enlargement of
blue and red bump spectral regions of galaxies where the broad
lines $HeII\lambda4686$ and CIV$\lambda5808$ are visible.}
\label{fig6.1}
\end{figure*}

\setcounter{figure}{6}
\begin{figure*}
\includegraphics[width=180mm,height=220mm]{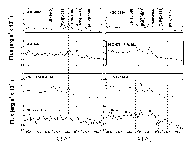} \caption{Continued.}
\label{fig6.2}
\end{figure*}

\setcounter{figure}{6}
\begin{figure*}
\includegraphics[width=180mm,height=220mm]{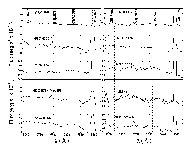} \caption{Continued.}
\label{fig6.3}
\end{figure*}

The dominant subtype of WR stars can be constrained by analyzing
the presence of particular lines. NIII$\lambda$4640 and/or
CIII$\lambda$4650 are observed in 9 galaxies, while
NV$\lambda$4604 is absent. The former are due only to WN stars,
thus indicating the predominance of WNL stars (Schaerer et al.
2000).

To derive the number of WNL stars from the blue bump, we measured
the flux of the entire bump, and then subtracted the nebular lines
and the contributions of NIII$\lambda$4640 and
CIII/CIV$\lambda$4650/$\lambda$4658 when these lines were present.
This procedure then retains only the broad emission component of
HeII$\lambda$4686.

The CIII$\lambda$5696 emission line is a signature of late type WC
stars  (WCL, GIT00), and their presence is expected in high
metallicity regions (Maeder, 1991). When CIII$\lambda$5696 is
absent, but CIV$\lambda$5808 is present, the emission is likely
due to early-type WC stars (WCE). WN stars cannot be responsible
for the CIV$\lambda$5808 emission, since the ratio
HeII$\lambda$4686/CIV$\lambda$5808 observed in our spectra ranges
from 0.64 to 5, much lower than the value of $\sim16$ predicted by
SV98 (Schaerer et al. 2000).

The absolute population of WR subtype stars can be derived if the
line luminosity of a single subtype star is known. We use the
integrated broad emission line luminosity of HeII$\lambda 4686$ in
each of our galaxies to derive the number of WNL stars, the
CIV$\lambda5808$ luminosity to derive the number of WCE stars, and
CIII$\lambda5696$ for WCL stars (GIT00).

In practice,
\begin{equation}
N_{WNL} = \frac{{L_{HeII\lambda 4686}^{obs} }}{{L_{HeII\lambda
4686} }},
\end{equation}
\begin{equation}
N_{WCE} = \frac{{L_{CIV\lambda 5808}^{obs} }}{{L_{CIV\lambda 5808}
}},
\end{equation}

\noindent and
\begin{equation}
N_{WCL} = \frac{{L_{CIII\lambda 5696}^{obs} }}{{L_{CIII\lambda
5696} }},
\end{equation}

\noindent where $L_{HeII\lambda4686}^{obs}$,
$L_{CIV\lambda5808}^{obs}$ and $L_{CIII\lambda5696}^{obs}$ are the
total luminosities observed around the corresponding spectral
features. For the average luminosity of a single WNL star in the
He II $\lambda $4686 line we use $L_{He\lambda4686} = 1.6\pm0.8
\times 10^{36}$ erg s$^{ - 1}$, while for a single WCE star in the
CIV$\lambda5808$ line we use $L_{CIV\lambda5808} = 3.0\pm1.1
\times 10^{36}$ erg s$^{ - 1}$, and for a single WCL star in the
CIII$\lambda5696$ line we use $L_{CIII\lambda5696} = 8.1\pm2.9
\times 10^{35}$ erg s$^{ - 1}$(SV98).

Then, the total number of WR stars is defined by
\begin{equation}
N_{WR}= N_{WCE} + N_{WCL} + N_{WNL}.
\end{equation}

We note that these line luminosities are based on WR stars observed in
the Milky Way, assuming solar metallicity . The line luminosities show
significant scatter depending on the dominant WR subtype and the
metallicity in the observed object.

To obtain the number of O stars in each galaxy, we
assume that all ionizing photons, $Q_{0}$, are produced by O and
WR stars (Conti 1991). Hence,
\begin{equation}
Q_0^{obs} = N_{O7V} Q_{O7V} + N_{WR} Q_{WR}
\end{equation}

\noindent where $N_{O7V}$ is the number of O7V stars and $Q_{WR}$ and
$Q_{O7V}$ are the number of ionizing photons per second produced
by WR stars (all subtypes summed up) and O7V stars, respectively.
Thus, the total number of ionizing photons can be obtained from
the H$\beta$ luminosity, $L(H\beta)$, through the relation

\begin{equation}
Q_0^{obs} = 2.01 \times 10^{12} L_{H\beta}.
\end{equation}

The number of O stars present is derived from the number of O7V
stars after applying a correction for the presence of other O star
subtypes. In this sense, VC92 and Vacca (1994) defined the
conversion parameter as the proportion of O7V stars relative to
all OV stars,

\begin{equation}
\eta _{0} = N_{O7V } / N_{OV }.
\end{equation}

The parameter $\eta_{0}$ depends on the IMF for massive stars and
is a function of the time elapsed since the beginning of the
burst.

Using the models of SV98, which give the evolution of EW(H$\beta$)
as a function of the time elapsed from the beginning of an
instantaneous burst for different metallicities, we derive the
starburst age $t$ for each of our galaxies. These ages are in good
agreement with the predicted age and duration of the WR phase
estimated from instantaneous burst models. We use the models of
SV98 to evaluate the parameter $\eta _{0}(t)$, adopting the oxygen
abundance obtained for objects as the metallicity, the canonical
slope for a Salpeter initial mass function $\Gamma = -2.35$, and a
stellar mass upper limit of $120 M_{\odot}$. The number of
ionizing photons, the age of the starburst and $\eta(t)$ derived
for each galaxy of our sample are given in Table 5.

\setcounter{table}{4}
\begin{table*}
\begin{minipage}{220mm}
\caption{Star Population Parameters.}
\begin{tabular}{lrrr}
\hline\hline

&  Q$_0$($\times10^{52}$)&  Age (Myr)& $\eta_{0}(t)$\\

\hline

Fairall44&        3.96$\pm$0.59&  3.0& 0.50\\

III Zw 107&      18.61$\pm$2.79&  4.0& 0.53\\

Mrk309&          89.76$\pm$13.46& 6.1& 0.80\\

Mrk475&           0.12$\pm$0.02&  4.1& 0.48\\

Mrk712&           2.27$\pm$0.34&  3.5& 0.86\\

Mrk1271&          0.04$\pm$0.01&  4.7& 0.36\\

NGC450&           3.52$\pm$0.53&  2.6& 1.50\\

NGC4385&          9.85$\pm$1.48&  4.9& 0.18\\

NGC4861&         3.18$\pm$0.48&  4.0& 0.53\\

NGC5430&        15.28$\pm$2.29&  5.1& 0.14\\

NGC5471&         0.03$\pm$0.00&  3.8& 0.65\\

NGC6764&        14.94$\pm$2.24&  5.5& 0.12\\

NGC6764&         12.40$\pm$1.86&  5.5& 0.12\\

NGC7714A&       29.68$\pm$4.45&  5.4& 0.52\\

NGC7714B&        2.79$\pm$0.42&  3.7& 0.72\\

UM48&            9.94$\pm$1.49&  4.8& 0.20\\

\hline
\end{tabular}

\end{minipage}
\end{table*}

Then, using equations (18),(19) and (20), the absolute number of O
stars can be derived as

\begin{equation}
 N_{O} = N_{OV} = \frac{{Q_0^{obs} - N_{WR} Q_{WR} }}{{\eta _0(t) Q_{O7V}
}}
\end{equation}

We adopt $Q_{WR}$ = $Q_{O7V} = 1.0 \times 10^{49} s^{ - 1}$
(Schaerer et al. 1999).

The absolute number of O stars, Wolf-Rayet stars, N$_{WNL}$ and
N$_{WCE}$, and the ratio N$_{WR}$/N$_{O}$ are given in Table 6.

\setcounter{table}{5}
\begin{table*}
\begin{minipage}{220mm}
\caption{Massive Star Population.}
\begin{tabular}{lrrrrrrrrr}
\hline\hline

&  {N$_{WNL}$}& N$_{WCL}$& N$_{WCE}$& N$_{WR}$&  N$_{O}$&
N$_{WR}$/N$_{O}$*&
N$_{WC}$/N$_{WN}*$\\

\hline

Fairall44&          692$\pm$363&              -&            -&   692$\pm$363&    5660$\pm$623&   0.12$\pm$0.04&              -\\

III Zw 107&       3389$\pm$2211&            -&            -&  3389$\pm$2211&  28726$\pm$4172&  0.13$\pm$0.04&              -\\

Mrk309&          30841$\pm$16314& 7945$\pm$229&            -& 38786$\pm$16315& 58498$\pm$16883& 0.66$\pm$0.24& 0.26$\pm$0.09\\

Mrk475&             25$\pm$14&              -&      7$\pm$3&    32$\pm$14&      186$\pm$29&    0.17$\pm$0.16& 0.28$\pm$0.10\\

Mrk712&            661$\pm$334&             -&            -&   661$\pm$334&    1876$\pm$390&   0.35$\pm$0.13&              -\\

Mrk1271&              11$\pm$7&               -&            -&    11$\pm$7&        77$\pm$22&    0.14$\pm$0.05&              -\\

NGC450&            322$\pm$215&             -&  129$\pm$51&    451$\pm$221&    2044$\pm$147&   0.22$\pm$0.08& 0.40$\pm$0.14\\

NGC4385&         3037$\pm$1534&   1208$\pm$960&  972$\pm$634&  5217$\pm$2288&  25740$\pm$7610&  0.20$\pm$0.07& 0.72$\pm$0.25\\

NGC4861&           522$\pm$288&      65$\pm$3&    58$\pm$35&    586$\pm$290&    4375$\pm$489&   0.15$\pm$0.05& 0.24$\pm$0.08\\

NGC5430&         4761$\pm$2520&            -&  999$\pm$435&  5761$\pm$2558&  66149$\pm$17776& 0.09$\pm$0.03& 0.21$\pm$0.08\\

NGC5471&               2$\pm$1&                -&            -&     2$\pm$1&        44$\pm$2&     0.05$\pm$0.02&              -\\

NGC6764&         2812$\pm$1687&     859$\pm$42& 1750$\pm$906&  5421$\pm$2231&  79325$\pm$14771& 0.07$\pm$0.02& 0.93$\pm$0.33\\

NGC6764&         2380$\pm$1457&     735$\pm$49& 1606$\pm$817&  4721$\pm$1943&  56180$\pm$11250& 0.08$\pm$0.03& 0.98$\pm$0.35\\

NGC7714A&        1920$\pm$1288&     869$\pm$18& 1596$\pm$883&  5295$\pm$2277&  44282$\pm$2655&  0.12$\pm$0.04& 0.87$\pm$0.31\\

NGC7714B&          451$\pm$255&               -&            -&   451$\pm$255&    3276$\pm$357&   0.14$\pm$0.04&              -\\

UM48&            1654$\pm$1279&             -&  353$\pm$234&  2007$\pm$1300&  38967$\pm$6385&  0.05$\pm$0.02& 0.21$\pm$0.08\\

\hline
\end{tabular}

\end{minipage}
\end{table*}

\subsection{Comparison with Evolutionary Models}

\label{section:comparationmodel}

The relationship between the N$_{WR}$/N$_{O}$ ratio and oxygen
abundance  obtained for the galaxies in our sample is shown in
Figure 8. We adopt a solar oxygen abundance of 12+ log(O/H) = 8.70
(Grevesse \& Anders 1989, Grevesse \& Sauval 1998). The
predictions of Schaerer $\&$ Vacca (1998) models for instantaneous
bursts (solid line) with IMF slopes of -1, -2 and  -2.35
(Salpeter), and of Starburst 99 (Leitherer et al. 1999) for
extended bursts (duration of 2-4 Myrs) with IMF  slope = -2.35 and
mass limit of 100 M$_{\odot}$ (dashed line) are plotted.

Evolutionary models predict that for a given metallicity, the
ratio between WR and O stars varies strongly with the age of the
burst, and the duration of the WR stage in the starburst also
increases with increasing metallicity. (Maeder \& Meynet 1994;
Meynet 1995; SV98). Maeder (1991) interpreted this behavior as the
result of increased stellar mass loss at higher metallicities. The
increased stellar mass loss reduces the mass limit
for forming WR stars in metal rich galaxies (Maeder 1991).

\setcounter{figure}{7}
\begin{figure*}
\includegraphics[width=150mm,height=220mm]{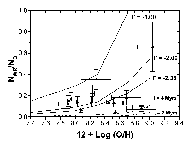} \caption{$N_{WR}/N_{O}$
versus $12+log(O/H)$ for the galaxies in our sample. Predictions
of Schaerer $\&$ Vacca (1998) models for instantaneous bursts
(solid line), and of Starburst 99 (Leitherer et al. 1999) for
extended burts of star formation of 2 and 4 Myrs. (dashed line)
are overplotted. The lines are labeled with the IMF slopes.}
\label{fig7}
\end{figure*}

In the low metallicity case, the results can be explained by an
instantaneous burst with a Salpeter IMF slope ($\Gamma$ = -2 to
-2.35). In the high metallicity regime, the results deviate from
the expected behavior based on models with an instantaneous burst
for starburst galaxies (SV98). A steeper IMF slope is required
than in the low metallicity case, or we must invoke an extended
burst to explain the results. SGIT00 interpret this behavior for
their sample of high metallicity galaxies as implying an extended
burst duration of $\sim$4-10 Myrs. Support for this conclusion
comes from the observed WR population relation (WC/WN) with
metallicity and the red supergiant star features observed in their
objects (GIT00,SGIT00). NGC4385, NGC 5430, NGC 6764 and UM 48,
with $12+log(O/H)>8.4$, are candidates for this type of object.

In Figure 9 we plot the ratio N$_{WC}$/N$_{WN}$ (where WC = WCL +
WCE) versus oxygen abundance compared to the predictions of SV98
models for an instantaneous burst, as well as extended bursts of 2
and 4 Myrs (Pindao et al. 2002). The dot-dashed line (bottom
right) shows the observed trend of WC/WN with metallicity in Local
Group galaxies derived empirically from observations by Massey
{\&} Johnson (1998). The N$_{WC}$/N$_{WN}$ ratio is lower than the
instantaneous burst model predictions and, for galaxies with
metallicity higher than $12+log(O/H)>8.4$, the ratio is closest to
the values reported for galaxies in the Local Group corresponding
to a constant star formation regime (Massey {\&} Johnson 1998).

These low N$_{WC}$/N$_{WN}$ ratios might be partially explained by
the assumption that the predominant contribution to the
CIV$\lambda$5808 line luminosity at high metallicities is from WCE
stars. If instead we assume that the main contribution to this
line luminosity comes from WCL stars, the lower luminosity of WCL
stars can increase the ratio of WC to WN stars by a factor of
$\sim3-4$ (GIT00). Schaerer {\&} Vacca 1998 predict a luminosity
ratio CIII$\lambda$5696/CIV$\lambda$5808 for a WCL (WC7) star of
$\sim 0.5$, but  galaxies in our sample where WC emission lines
are measured show lower CIII$\lambda$5696/CIV$\lambda$5808 ratios.

We contend that the low N$_{WC}$/N$_{WN}$ ratios observed in high
metallicity galaxies are due to the nature of the star formation
bursts, with different durations for the WN and WC stages (GIT00).
These galaxies have low H$\beta$ equivalent widths leading to ages
greater than 5.3 Myr (except UM48 with age $\simeq 4.8$ Myr). They
are therefore in the late stages of their WR episodes (GIT00). At
these later times, WN stars are still present, while the number of
WC star drops to zero (SV98). This supports an extended burst to
explain the N$_{WN}$/N$_O$ ratio seen in these galaxies.

\setcounter{figure}{8}
\begin{figure*}
\includegraphics[width=150mm,height=220mm]{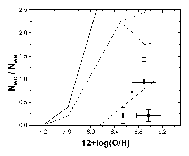} \caption{$N_{WC}/N_{WN}$
versus $12+log(O/H)$ for the high metallicity galaxies in our
sample. Predictions of Schaerer $\&$ Vacca (1998) models for
maximum value for instantaneous bursts (solid line), and SGIT00
extended bursts of 2 - 4 Myrs (dashed lines) are overplotted. The
dot-dashed line (bottom right) shows the observed trend of WC/WN
with metallicity in Local Group galaxies derived empirically from
observations by Massey {\&} Johnson (1998).} \label{fig9}
\end{figure*}

The presence of late type stellar features provides additional
evidence supporting an extended burst.  These absorption features
are commonly observed in integrated spectra of stellar clusters,
indicating the presence of red giants and supergiants (Bica {\&}
Alloin 1986). The presence of late-type stars further supports the
idea of an extended burst with an age $<$7 Myr. In Figure 10 we
present the spectra of these four high-metallicity galaxies which
show apparent TiO bands ($\sim \lambda6250$) and and a blend of
FeI+BaII+CaI$\lambda$6495, which are spectral features of
characteristic of late-type stars.

\setcounter{figure}{9}
\begin{figure*}
\includegraphics[width=150mm,height=180mm]{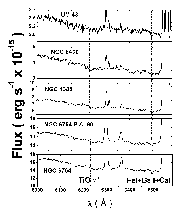} \caption{Spectral region
showing TiO bands ($\sim \lambda6250$) and narrow FeI+BaII+CaI
$\lambda$6495 for the high-metallicity galaxies NGC 4385, NGC
5430, NGC 6764 (P.A. 67 and 90) NGC 5471 and UM 48.} \label{fig8}
\end{figure*}

Alternatively, Pindao et al. (2002) suggest that corrections to
the evolutionary models can reproduce the low observed
N$_{WR}$/N$_{O}$ ratios using an instantaneous burst. They invoke
the uncertainties in synthesis models of WR bumps to explain their
results, since the WR line luminosities show a large scatter in
the WNL calibration sample of SV98 with Galactic and LMC objects.
A dependence between the WR luminosity and the bolometric stellar
luminosity may cause this discrepancy. The WR line luminosity
L$_{\rm4686}$ is observed to increase with increasing bolometric
stellar luminosity. They suggest a more precise analysis,
splitting the bolometric stellar luminosity in two domains
($log(L/L_{\odot})<6$ and $log(L/L_{\odot})>6$). The different
average line luminosities for each domain are L$_{4686} =
5.6\times 10^{35}$ erg s$^{ - 1}$ for $log(L)<6$ and L$_{4686} =
3.1\times 10^{36}$ erg s$^{ - 1}$ for $log(L)>6$.  Using these
values in synthesis models leads to a reduction in the predicted
number of WNL stars. Note that the line luminosity for a WNL star
with $L>6L_{\odot}$ is almost twice as large as that adopted by
SV98 for their Galactic calibration sample. Using these values in
the synthesis models, their observed WR features can be reproduced
with an instantaneous burst and a standard Salpeter IMF (Pindao et
al. 2002).The recalibrated (lower) luminosity of WN stars used in
models leads to a significant reduction in the WR bump intensity
in the bursts with ages $\geq$ 4-5 Myrs.

Only the youngest bursts with very high $EW(H\beta)$ are dominated
by very luminous WNL stars. For objects with low $N_{WR}/N_{O}$
ratios, if we apply the suggested correction to the L$_{\rm4686}$
luminosities in the actual observations (instead of the models),
and recalculate the number of WN stars at the highest
metallicities (dominated by low luminosity WNL stars), the lower
luminosity of WNL stars can increase the fraction of WN stars by a
factor of $\sim 2-2.3$ (GIT00). With this change in assumed
luminosity, we are unable to reproduce the standard instantaneous
burst model predictions for the $N_{WR}/N_{O}$ ratio in the
highest metallicity galaxies. This analysis confirms that the low
values of N$_{WR}$/N$_{O}$ for objects with $12+log(O/H)>8.4$,
obtained for NGC4385, NGC 5430, NGC 6764 and UM 48, require an
extended burst with a Salpeter IMF slope.

\section{Conclusion}
\label{section:conclusion}

In this work, we present a spectroscopic study of 14 Wolf-Rayet
galaxies from the sample of SCP98, as well as NGC450 for which the
WR features are newly detected. Our goals were to search for and
confirm the presence of WN and WC stars, and to compare the
results with predictions from evolutionary synthesis models
(SV98). We tested the agreement of these models with observations
for a large range of metallicities, spanning $7.90\le$ 12 +
log(O/H) $\le 9.04$.

Our main results can be summarized as follows:

1. The broad WR emission in the blue region of the spectrum, the
blend of NIII$\lambda 4640$, CIII$\lambda4650$, CIV$\lambda4658$,
and HeII$\lambda4686$ emission lines, is present in all fourteen
galaxies. The WR population in these galaxies is dominated by late
WN stars. However, the red bump produced by the emission of broad
C IV$\lambda5808$ from early WC stars is detected in only nine
galaxies.

2. The weak, broad WR emission line CIII$\lambda5696$ is detected
in six galaxies, which suggests the presence of late WCL stars in
these objects. This line is expected in high-metallicity
environments (GIT2000). A good example of this is Mrk 309, with a
metallicity of 12+log(O/H) = 9.04, where this line is strong.

3. We found good agreement when comparing the relative numbers of
WR and O stars ($N_{WR}/N_{O}$) obtained from observations and
those predicted by the evolutionary synthesis models of SV98 for
low metallicity galaxies.  The ratio $N_{WR}/N_{O}$ in these
galaxies can be explained by a burst of star formation. The
$N_{WR}/N_{O}$ value and the observed equivalent widths of the
blue and red bumps also compare favorably with SV98 predictions.
We found that it is necessary to invoke an IMF slope between
$-2\lsim\Gamma\lsim-2.35$ and an instantaneous star formation
event to explain the observed $N_{WR}/N_{O}$ ratios in low
metallicity regimes.

4. For NGC4385, NGC 5430, NGC 6764 and UM 48, the $N_{WR}/N_{O}$
ratios are lower than the predictions of models with an
instantaneous burst for metallicities of $12+log(O/H)>8.4$. The
existence of an extended burst is supported by the presence of TiO
bands ($\sim \lambda$6250) in these objects spectra. The presence
of this older stellar population is indicative of the necessary
time elapsed since the burst for stars to evolve through this
phase. The $N_{WN}/N_{WC}$ ratios combined with their high
metallicity suggest these objects are in the late stages of their
WR episodes ($\geq 5.3Myr$), when WN stars are present while the
number of WC stars drops to zero (SV98). The observations can be
better represented using extended starbursts with durations of 2-4
Myr and Salpeter IMF slope.

5. Massive stellar evolution models predict that the relative
number of WR stars increases when metallicity increases (Maeder \&
Meynet 1994, SV98). Our results do not confirm this trend. The
partial disagreement between our results and the models may arise
from large uncertainties in the luminosity of a single WR star and
the uncertainties associated with the best choice of the dominant
contribution of WR star subtype in the high metallicity regime.

The massive stellar populations in Wolf-Rayet galaxies can be
better understood if studies like this one are combined with other
stellar population investigations. A detailed spectral study in
the infrared of molecular bands and low ionization absorption
lines can provide information about older stellar populations in a
starburst (Origlia et al. 1999, GIT00). Spectroscopy in the UV
($912- 1800\AA$) can provide information about the young star
population in galaxies. This UV spectral range contains resonant
spectral lines of OVI$\lambda1035$, SiV$\lambda1400$ and/or
CIV$\lambda1550$ that are spectral signatures of young massive
stars (Leitherer et al. 2002). The combination of the IR and UV
yields information on the current stellar populations, and
constrains the upper and lower limits of time elapsed since the
burst (Leitherer et al. 2002, Olivia et al. 1999). Additionally,
analysis of the $\alpha$-element abundances can provide additional
constraints on the age of the burst and consequently the number of
massive stars (Lanfranchi, G. A. {\&} Fria\c ca 2003).

$Acknowledgements:$ It is a pleasure to thank Sueli M. M. Viegas
for discussions and contributions with helpful comments.
I.F.Fernandes is thankful to the Laboratoire d'Astrophysique,
Observatoire Midi-Pyr\'{e}n\'{e}es staff for their kind
hospitality. This international collaboration was possible thanks
to the financial support of IAG-USP, FAPESP grant No 99/12721-5
and of . L.A.O.M.P. URA 285.

\section*{Appendix A: Remarks on Individual Objects}
\label{section:appendixB}

In this section we present a brief description of some specific
properties of each galaxy in our sample. By doing so we can
address specific issues related to these objects and compare with
results from other authors, and obtain important information for
the analysis of the population as a whole.

Fairall 44: Kovo \& Contini (1999) reported the blue bump in their
systematic search for Wolf-Rayet stars in young starburst
galaxies. We report the presence of WN stars in this galaxy. We
find that the WR/O ratio is 0.24.

III Zw 107: Kunth \& Jobert (1985) reported a moderately strong
emission band at $4686$\AA~ due to WR stars in III Zw 107 S. We
report the presence of NIII$\lambda4640$ and HeII$\lambda4886$
broad emission lines in this galaxy. The WR/O ratio is 0.12.

Mrk 309: This bright UV continuum galaxy has broad emission
features at NIII$\lambda4640$ and HeII$\lambda4686$ from WR stars
noted by OC82. They found that the number of WR stars is
comparable to the number of O stars in this galaxy and the nuclear
region of NGC6764. We note the presence of NIII$\lambda4640$ and
HeII$\lambda4686$ in the blue bump and CIII$\lambda5696$ in the
red bump in our spectrum of this galaxy. We find that the number
of WR stars is similar to that of O type stars in the nuclear
region of Mrk 309. Our results are in good agreement with SGIT00.

Mrk 475: Broad HeII$\lambda4686$ and NIII$\lambda4640$ emission
lines were first noted by Conti (1991). Strong blue and red bumps
were detected by ITL94. GIT00 identified features of
SiIII$\lambda4565$, HeII$\lambda4686$, HeI,NII$\lambda5047$, and
CIV$\lambda5808$, where the blue bump is strongly contaminated by
nebular emission. We confirm the results of GIT00, and find a WR/O
ratio of 0.17.

Mrk 712: Contini et al. (1995) reported the discovery of emission
from WR stars in the giant HII region $4.5^{\prime\prime}$ south
of the nucleus. They estimated a WN/O ratio of 0.2 from the
HeII$\lambda4686$ luminosity. The [ArV] emission line in their
spectrum showed that the HII region is strongly ionized by hot WR
stars. The WN/WO ratio indicated a very young starburst episode
and a flat initial mass function with slope $\Gamma$ between -1
and -2. We use the SV98 evolution models to compare with our
observational results (\S6.2).

Mrk 1271: Izotov \& Thuan 1998 detected a broad blue bump. GIT00
identified the NIII$\lambda4512$, SiIII$\lambda4565$,
NII$\lambda4620$ and HeII$\lambda4686$ emission lines, and
possibly CIV$\lambda5808$ in their high quality spectrum. They
reported the blue bump as being strongly contaminated by nebular
emission. We confirm only the presence of the HeII$\lambda4686$
broad line in the nuclear region of this galaxy. The
CIV$\lambda5808$ broad line is not detected in our spectrum of Mrk
1271. The WR/O ratio is 0.14.

NGC 450: There is no mention of WR detection in this galaxy in the
literature. We report a ratio of WR/O=0.22. The models of SV98 for
an instantaneous star formation episode predict a flat IMF slope
($\Gamma= -1$ to $-2$) to explain the presence of WR stars in this
low metallicity galaxy.

NGC 4385: HeII$\lambda4686$ and NIII$\lambda4640$ Wolf-Rayet
features are present in this starburst galaxy, according to Durret
\& Tarrab (1988). An optical spectrum by Salzer (1990) shows
essentially the same features. Conti (1991) identified a narrow
emission feature near 4660\AA~ as [FeIII], while Salzer et al.
(1989) identifies it as CIV. We detect NIII$\lambda4640$ and
HeII$\lambda4686$ in the blue bump and CIII$\lambda5696$ and
CIV$\lambda 5808$ in the red bump. We find a high number of WR
stars compared to O stars. The WR/O ratio is 0.13 for the nuclear
region of NGC 4385.

NGC 4861: Dinerstein \& Shields (1986) and Izotov, Thuan \&
Lipovetsky (1997) detected the blue and red bumps. GIT00
identified NIII$\lambda4512$, SiIII$\lambda4565$, $NV\lambda4619$,
HeII$\lambda4686$ and CIV$\lambda5808$ broad emission lines, and a
blue bump strongly contaminated by nebular emission. Even with
their high-quality spectrum, the NIII$\lambda4640$ and
CIV$\lambda4658$ lines appear blended. In our spectrum of the
central region of this galaxy, we can resolve the lines of the
blue bump. We report the presence of CIII$\lambda5696$ and
CIV$\lambda5808$ broad emission lines due to WC stars.

NGC 5430: A strong emission feature near 4650\AA~ due to WR stars
in the spectrum of a bright region SE of the galactic center was
noted by Kell (1982). He identified NIII$\lambda4640$ and
HeII$\lambda4686$ emission lines as coming from WN stars (Kell
1987). No strong emission features are seen in the IUE spectrum of
this source. A relatively older stellar population appears in the
center of NGC 5430. According to Kell (1987), the knot
20$^{\prime\prime}$ southeast of the nucleus might be a separate
galaxy interacting with NGC 5430. For this knot we measure the
NIII$\lambda4640$ and HeII$\lambda4686$ blue bump broad
components, as well as the CIII$\lambda5696$ and CIV$\lambda5808$
broad emission lines. A high WR/O ratio of 0.09 is found for NGC
5430.

NGC 5471: This is a massive giant HII region located in M101.
Mass-Hesse \& Kunth (1991) found WR-HeII$\lambda4686/$H$\beta=
0.02$ and a 3.5 Myr burst. NGC 5471 seems to be dominated by a
well-defined burst of star formation. A broad HeII emission line
(FWHM$\simeq$2000 km s$^{-1}$) was detected in this region by
Casta\~{n}eda et al. (1990) and was confirmed by Mass-Hesse et al.
(1991). We find a low ratio WR/O=0.04 for this galaxy.

NGC 6764: OC82 noted the presence of broad NIII$\lambda4640$ and
HeII$\lambda4686$ emission line features from WR stars. They
attributed the 4660\AA~ emission line to CIII and not to [FeIII]
forbidden emission. The similarity of the overall emission line
spectrum of NGC 6764 to that of dwarf galaxies and to certain
giant HII regions attracted the attention of these authors. We
observed this object at two position angles, $90^{\circ}$ and
$46^{\circ}$. We find WN and WC stars in the nuclear region of
this galaxy. For both position angles the number of Wolf Rayet
stars is comparable to that of O type stars.

NGC 7714: Weedman et al. (1981) called this object a ``prototype
starburst'' galaxy. van Breugel et al. (1985) noted the spectral
similarity of NGC7714 to Minkowski's object and extragalactic HII
regions. They reported this galaxy as having weak WR emission
features near HeII$\lambda4686$. Conti (1991) called attention to
the importance of NGC 7714 for understanding the relation between
starbursts and the presence of WR stars in this kind of galaxy. We
found WR stars in two regions of NGC 7714. In the nuclear region
we find WR/O = 0.08, while in the secondary HII region, WR/O=0.14.
In the nuclear regions we were able to measure the CIII$\lambda
5696$ and CIV$\lambda5808$ broad emission lines due to WC stars.

UM48: A systematic search for Wolf-Rayet features in this galaxy
was done by Masegosa et al. (1991). The authors reported the
presence of a WR blue bump. They assumed that the global WR
detection rate depends on the metallicity and they analyzed the SN
IIe contribution to the blue bump. They concluded that a large
blue bump luminosity should be expected in regions with SN
contamination. We did not detect the NIII$\lambda4640$ broad
emission line in this galaxy. We measure the WC broad emission
line CIV$\lambda5808$.

\bsp \label{lastpage}
\end{document}